\documentclass[twocolumn,showpacs]{revtex4}

\usepackage{graphicx}
\usepackage{dcolumn}
\usepackage{bm}

\begin{document}

\title{Spin-gap phase of a quantum spin system 
on a honeycomb lattice}

\author{Ken'ichi Takano}
\affiliation{Toyota Technological Institute, 
Tenpaku-ku, Nagoya 468-8511, Japan}


\begin{abstract}

	We study a quantum spin system on a honeycomb lattice, 
when it includes frustration and distortion 
in antiferromagnetic (AF) exchange interactions. 
	We transform the spin system onto a nonlinear $\sigma$ 
model (NLSM) in a new way preserving the original spin 
degrees of freedom. 
	Assisted by a renormalization-group argument, the NLSM 
provides a ground-state phase diagram consisting of 
an ordered AF phase and a disordered spin-gap phase. 
	The spin-gap phase extends from a strong frustration 
regime to a strong distortion regime, showing that 
the disordered ground states are essentially the same 
in both the regimes. 
	In the spin-half case, the spin-gap phase for the spin system 
on a honeycomb lattice is larger than that for 
the $J_1$-$J_2$ model on a square lattice. 

\end{abstract}    


\maketitle


	Recently, materials regarded as spin systems on honeycomb 
lattices have been found. 
	Kataev et al.~\cite{Kataev} synthesized 
a spin-half ($S=\frac{1}{2}$) material on a honeycomb lattice, 
InCu${}_{2/3}$V${}_{1/3}$O$_3$. 
	They estimated the first-neighbor exchange parameter as 
280 K, and found an antiferromagnetic (AF) ordering at 38K 
with interlayer interactions. 
	Their results are confirmed in refined samples 
by Kikuchi et al.~\cite{Kikuchi}. 
	Miura et al.~\cite{Miura} reported other materials, 
Na$_3$T$_2$SbO$_6$ (T = Cu, Ni, and Co). 
	The material with T = Cu is an $S=\frac{1}{2}$ system and 
shows a spin-gap in magnetic susceptibility. 
	The materials with T = Ni and Co are 
$S=1$ and $S=\frac{3}{2}$ systems, respectively, 
and both have AF order. 
	So far, samples are reported to have fairly strong 
dimer-like distortion (T = Cu)~\cite{fitting} or randomness 
(T = Ni and Co) in exchange interactions on honeycomb lattices. 
	In view of recent experimental activity, 
materials regarded as various types of spin systems 
on honeycomb lattices are expected to be synthesized. 
	Hence it is worthwhile to show a theoretical scheme for 
spin systems on honeycomb lattices. 

	In two dimensions,  there are two kinds of simple 
bipartite lattices, a square lattice and a honeycomb lattice, 
if only first-neighbor AF interactions are considered. 
	A spin system on a bipartite lattice has a ground state 
with an AF order in a large spin-magnitude limit. 
	A question is whether or not a disordered spin-gap phase 
appears with destroyed AF order, when the spin magnitude 
becomes small so that quantum fluctuations are large. 
	Quantum fluctuations are further enhanced by frustration 
among AF interactions 
and by distortion as inhomogeneity in interactions. 
	Hence a disordered state will be interestingly studied 
under variable strengths of frustration and distortion.  

	In the square-lattice case, 
the $J_1$-$J_2$ model has been eagerly studied. 
	Theoretically, whether or not a disordered state is formed 
around $J_2/J_1 = 0.5$ is a problem 
controversial for long time. 
	The difficulty originates from the fact that 
the region of the disordered phase is very narrow  
in the $J_1$-$J_2$ parameter space even if it exists. 
	Further a material realizing the $J_1$-$J_2$ model 
in a disordered state has not been found. 
	The difficult synthesis of such a material also seems to 
come from the narrowness of the disordered phase. 

	In contrast, a quantum spin system on a honeycomb lattice 
is more hopeful to find a disordered ground state.
	In fact, the coordination number 3 of a honeycomb lattice 
is smaller than 4 of a square lattice, and so the AF order 
for a honeycomb lattice is more fragile. 
	The case without distortion is treated by a numerical 
diagonalization and a linear spin-wave theory~\cite{Fouet}, 
Schwinger-boson mean-field theory~\cite{Mattsson}, and 
a nonlinear $\sigma$ model (NLSM) method~\cite{Einarsson}. 
	A systematic study of a spin system on a honeycomb lattice 
both with frustration and distortion is expected. 

	Among various methods to analyze a two-dimensional 
spin system, mapping it onto an NLSM and 
applying a renormalization group (RG) analysis is 
effective to study a quantum phase transition. 
	In the case of a square lattice, 
a spin system only with first-neighbor exchange interactions 
is shown to be mapped onto an NLSM without topological 
term~\cite{nontopological}. 
	For a $J_1$-$J_2$ model, NLSM methods are developed 
and a spin-gap phase are studied~\cite{Chakravarty,Einarsson}.  
	After then, a refined mapping onto an NLSM is proposed
for a $J_1$-$J_2$ model on a square lattice 
with plaquette distortion~\cite{Takano}. 
	In the derivation, the original spin degrees of freedom 
are properly treated without doubling by a variable 
transformation, in which a slowly varying AF variable appears 
for a block of four spins. 
	The derived NLSM has no topological term even if 
a second-neighbor interaction and a plaquette distortion exist. 
	The RG analysis~\cite{Chakravarty} is successfully applied 
to the obtained NLSM and provides a phase diagram 
in the space of parameters describing frustration and 
plaquette distortion~\cite{Takano}. 

	In this paper, we propose a novel NLSM 
method for a spin 
system on a honeycomb lattice in the case with frustration 
by second-neighbor AF interactions and dimer-like distortion 
in first-neighbor AF interactions. 
	The original spin Hamiltonian is mapped onto a field theory 
in a manner that six close spins are transformed into 
one slow modulation of the AF spin configuration and 
five rapid motions. 
	The degrees of freedom of the original spin model are 
properly considered through the derivation, and the choice of 
the cutoff becomes unique~\cite{Einarsson2}. 
	The derived field theory is an NLSM except for 
an additional term. 
	After dealing with the additional term, we apply the RG 
analysis~\cite{Chakravarty} to the NLSM to obtain 
a ground-state phase diagram; 
	We obtain a ground-state phase diagram 
in the parameter space, and show a disordered region 
in which there is a spin-gap. 

\begin{figure}[btp]
\begin{center}\leavevmode
\includegraphics[width=0.75\linewidth]{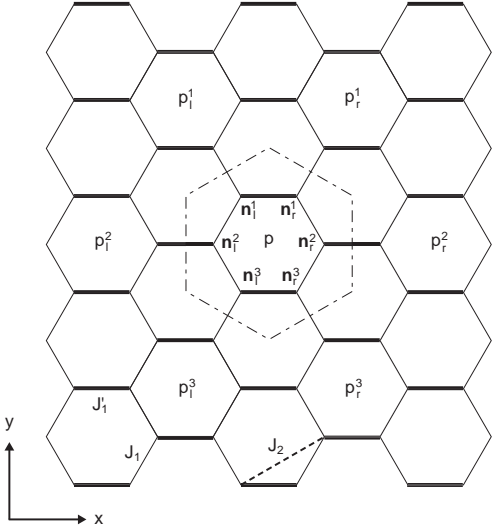}
\end{center}
\caption{Spin system on a honeycomb lattice, where 
a spin with magnitude $S$ is on a lattice point. 
	A first-neighbor exchange interaction with coupling constant 
$J_1$ ($J'_1$) is represented by thin (bold) solid line.  
	A second-neighbor exchange interaction with coupling 
constant $J_2$ is represented by a dotted line; 
dotted lines should be between any second-neighbor 
spins, although not drawn to avoid complication. 
	The $p$th block is shown by a dash-dotted line; 
the adjacent blocks are labeled as $p^{j}_{u}$ 
$(j = 1, 2, 3; u = l, r)$. 
	Variables ${\bf n}^{j}_{u}$ $(j = 1, 2, 3; u = l, r)$ represent 
spins in the path-integral formula with spin coherent states. 
} 
\label{lattice}
\end{figure}

      The Hamiltonian of the present spin system 
on a honeycomb lattice is  
\begin{equation}
H = \sum_{\langle i, j \rangle} 
J_{ij} {\bf S}_i \cdot {\bf S}_j , 
\label{Hamiltonian}
\end{equation}
where ${\bm S}_i$ is the quantum spin at site $i$ and 
the spin magnitude $S$ is uniform and arbitrary. 
      Exchange constant $J_{ij}$ takes 
$J_1$, $J'_1$ or $J_2$ depending on the bond between 
sites $i$ and $j$ as shown in Fig.~\ref{lattice}. 
	The distance between first-neighbor spins will be 
written as $a$. 
      We consider the quantum Hamiltonian 
in the classical N\'eel ordered region. 

	Quantum fluctuation is increased by distortion and 
frustration in exchange constants. 
	To measure the strengths, we  introduce 
distortion parameter $\delta$ and 
frustration parameter $\alpha$ as follows:  
\begin{eqnarray}
J_1 = {\bar J}_1 (1 -  \delta) , \ 
J'_1 = {\bar J}_1 (1 + 2\delta) , \ 
\alpha = J_2/{\bar J}_1 . 
\label{parameter}
\end{eqnarray}
	Then ${\bar J}_1$ is an average of all first-neighbor 
exchange constants: ${\bar J}_1 = \frac{1}{3} (2J_1 + J'_1)$. 

	We construct a path-integral formula by means of 
spin coherent states. 
	For a spin coherent state labeled by a unit vector 
${\bf n}_j(\tau)$  at imaginary time $\tau$, 
the expectation value of ${\bf S}_j$ is represented  as 
$\langle {\bf S}_j \rangle = \zeta_j S {\bf n}_j(\tau)$, 
where $\zeta_j$ takes $+$ or $-$ depending 
on the sublattice which the $j$th site belongs to. 
	Then the general formula of the partition function 
at temperature $1/\beta$ is written as 
\begin{equation}
           Z =  \! \int \!\! D[{\bf n}_j(\tau)] \, 
\prod_j \delta(({\bf n}_j(\tau))^2 - 1) \, e^{-A}  
\label{partition}
\end{equation}
with action 
\begin{equation}
      A = i S \sum_{j} \zeta_j w[{\bf n}_j] 
              + \int_0^{\beta}  \! d\tau H(\tau) . 
\label{action_n} 
\end{equation}
      The first term in $A$ is the Berry phase term 
with the solid angle $w[{\bf n}_j]$ which the unit 
vector ${\bf n}_j(\tau)$ forms in period $\beta$. 
	Hamiltonian $H(\tau)$ is written as 
\begin{eqnarray}
H(\tau) = \frac{1}{2} S^2 
\sum_{\langle i, j \rangle} 
J_{ij} [{\bf n}_i(\tau) - {\bf n}_j(\tau)]^2 , 
\label{Hamiltonian_n}
\end{eqnarray}
using constraint $({\bf n}_j(\tau))^2 = 1$. 
      Hereafter we do not explicitly denote the $\tau$ 
dependence of ${\bf n}_j(\tau)$. 

	By finding an appropriate variable transformation, 
we will separate out a slow mode describing 
AF fluctuation for the honeycomb lattice. 
	We adopt the region enclosed by a dash-dotted line in 
Fig.~\ref{lattice} as a unit of transformation, 
and call it a {\it block}\,; 
the first-neighbor blocks to the $p$th block are labeled  
by $p^{j}_{u}$ ($j$ = 1, 2, 3 and $u = r, l$). 
	Accordingly we relabel the six variables, ${\bf n}_{j}$'s, 
in the $p$th block as ${\bf n}_{r}^{j}(p)$ and 
${\bf n}_{l}^{j}(p)$ ($j$ = 1, 2, 3), as shown in Fig. \ref{lattice}. 
	Then we perform the following variable transformation 
for each block:  
\begin{eqnarray}
{\bf n}_{r}^{1}(p) &=& {\bf m}(p)  
+ 3a[ - {\bf L}(p) + {\bf X}_{-}(p)  +  {\bf Y}_{-}(p) ] , 
\nonumber \\
{\bf n}_{r}^{2}(p) &=& {\bf m}(p)  
+ 3a[ {\bf L}(p) + {\bf X}_{+}(p)  +  {\bf Y}_{+}(p) ] , 
\nonumber \\
{\bf n}_{r}^{3}(p) &=& {\bf m}(p)  
+ 3a[ - {\bf L}(p) - {\bf X}_{-}(p)  -  {\bf Y}_{-}(p) ] , 
\nonumber \\
{\bf n}_{l}^{1}(p) &=& {\bf m}(p)  
+ 3a[ {\bf L}(p) + {\bf X}_{-}(p)  -  {\bf Y}_{-}(p) ] , 
\nonumber \\
{\bf n}_{l}^{2}(p) &=& {\bf m}(p)  
+ 3a[ - {\bf L}(p) - {\bf X}_{+}(p)  +  {\bf Y}_{+}(p) ] , 
\nonumber \\
{\bf n}_{l}^{3}(p) &=& {\bf m}(p)  
+ 3a[ {\bf L}(p) - {\bf X}_{-}(p)  +  {\bf Y}_{-}(p) ] . 
\label{transform}
\end{eqnarray}
	Here ${\bf L}(p)$, ${\bf X}_{\pm}(p)$, and 
${\bf Y}_{\pm}(p)$ describe fluctuation modes 
around AF variable ${\bf m}(p)$. 
      Six original constraints, 
$({\bf n}_{t}^{j}(p))^2$ = 1 ($j$ = 1, 2, 3 and $t$ = $r$, $l$), 
are changed to six new constraints, $({\bf m}(p))^2$=1 and 
${\bf m}(p)$$\cdot$${\bf L}(p)$ = 0, 
${\bf m}(p)$$\cdot$${\bf X}_{\pm}(p)$ = 0, and 
${\bf m}(p)$$\cdot$${\bf Y}_{\pm}(p)$ = 0~\cite{normalization}. 

	Taking a continuum limit, the first term 
of action (\ref{action_n}) is written as  
\begin{eqnarray}
&& iS \sum_p \sum_{j=1}^3 
    (-1)^j \{ w[{\bf n}_r^{j}(p)] - w[{\bf n}_l^{j}(p)] \} 
\nonumber \\ 
&& \rightarrow  -i \frac{2S}{3a} \int \!\! d\tau d^2{\bf r} \, 
         ( 3 {\bf L} + {\bf X}_{+} )
\! \cdot \! ({\bf m} \! \times \! \partial_{\tau} {\bf m}). 
\end{eqnarray}
	In each term of Eq.~(\ref{Hamiltonian_n}), 
calculation like the following is carried out; 
\begin{eqnarray}
&& {\bf n}_{r}^{2}(p) - {\bf n}_{l}^{2}(p_{r}^{2}) = 
{\bf m}(p) - {\bf m}(p_{r}^{2}) + 6a({\bf L} + {\bf X}_+) 
\nonumber \\ 
&& \rightarrow 3a [ \partial_x{\bf m} 
+ 2({\bf L} + {\bf X}_+) ] 
\end{eqnarray}
up to the lowest order of derivatives and fluctuation variables. 
	Thus the lattice action (\ref{action_n}) becomes to 
the field-theoretic action: 
\begin{eqnarray}
A &=& S^2 \! \int \!\!\!\! \int \!\!\!\! \int \! d{\tau}d^2{\bf r} 
\, \Bigl\{ \, 
\frac{3}{4}({\bar J}'_1 - 4J_2) (\partial_x{\bf m})^2 
\nonumber \\ 
&+& \frac{3}{4}(J_1 - 4J_2) (\partial_y{\bf m})^2 
\nonumber \\ 
&+& 18{\bar J}_1 {\bf L}^2 
+ 6{\bf L}\cdot(2{\bar J}_1{\bf X}_{+} + {\bf B}_{-})
\nonumber \\ 
&+& 2(J_1 + J'_1 - 3J_2) {\bf X}_{+}^2 
       - 2{\bf X}_{+}\cdot({\bf B}_{+}  -3J_2 \partial_x{\bf m})
\nonumber \\ 
&+& 2(J_1 - 3J_2) ( 3{\bf X}_{-}^2 
   - \sqrt{3}{\bf X}_{-}\cdot \partial_y{\bf m}+{\bf Y}_{+}^2 ) 
\nonumber \\ 
&+& 6({\bar J}'_1 - 3J_2) {\bf Y}_{-}^2 \Bigr\} 
\label{action_mL} 
\end{eqnarray}
with 
${\bf B}_{+} = \frac{1}{3} [ 3J'_1 \, \partial_x{\bf m} 
+ i (aS)^{-1} \, {\bf m} \! \times \! \partial_{\tau} {\bf m} ]$, 
${\bf B}_{-} = \frac{1}{3} [ (J_1~-~J'_1) \, \partial_x{\bf m}  
- i (aS)^{-1} \, {\bf m} \! \times \! \partial_{\tau} {\bf m} ]$, 
${\bar J}_1 = \frac{1}{3}(2J_1 + J'_1)$, 
and ${\bar J}'_1 = \frac{1}{3}(J_1 + 2J'_1)$. 
      This action includes all the low-energy 
excitations surviving the continuum approximation, 
since the original degrees of freedom are not spoiled 
in the variable transformation (\ref{transform}). 

      Integrating out the partition function with 
the action (\ref{action_mL}) with respect to 
massive fields ${\bf L}$, ${\bf X}_{\pm}$, and ${\bf Y}_{\pm}$, 
we obtain the following action: 
\begin{eqnarray}
&& A'  = \int \!\!\!\! \int \!\!\!\! \int \! d\tau d^2{\bf r} 
\biggl\{ -i \frac{S\delta}{3a} \, {\bf m} \cdot 
(\partial_{\tau} {\bf m} \! \times \! \partial_{x} {\bf m}) 
\nonumber \\ 
&& + \frac{1}{18a^2{\bar J}_1} (\partial_{\tau}{\bf m})^2 
+ \frac{1}{4} S^2{\bar J}_1 
[ c_+ (\partial_x{\bf m})^2 
+ c_-(\partial_y{\bf m})^2 ] 
\biggl\} 
\nonumber \\ 
\label{action_prime} 
\end{eqnarray}
with $c_+ = 1+ \delta - 2\delta^2 -6\alpha$ 
and $c_- = 1- \delta - 6\alpha$. 
	Except for the first term this is an NLSM action. 
	The first term is rewritten as 
\begin{eqnarray}
-iA_{Q} = -i \frac{S\delta}{3a} \int \! dy Q(y) 
\label{first_term} 
\end{eqnarray}
with 
\begin{eqnarray}
Q(y) = \int \!\!\!\! \int \! d\tau dx \ 
{\bf m} \cdot (\partial_{\tau} {\bf m} \! \times \! \partial_{x} {\bf m}). 
\label{charge} 
\end{eqnarray}
	The quantity $Q(y)$ is not a topological charge 
in the three-dimensional Euclidean space. 
	For a fixed value of $y$, however, $Q(y)$ is the topological 
charge of the field ${\bf m}$ in the Euclidean $\tau$-$x$ plane, 
and is an integer. 
	Further, for the continuity, $Q(y)$ must be the same 
integer for all values of $y$. 
	This means that there is at least a vortex line for $Q \ne 0$. 
	Hence, a configuration with $Q \ne 0$ gives positive 
contributions of the system size to the second and 
the third terms in the action (\ref{action_prime}). 
	Thus we remove the first term from the action 
(\ref{action_prime}), assuming that 
configurations with $Q \ne 0$ are negligible. 

\begin{figure}[btp]
\begin{center}\leavevmode
\includegraphics[width=0.9\linewidth]{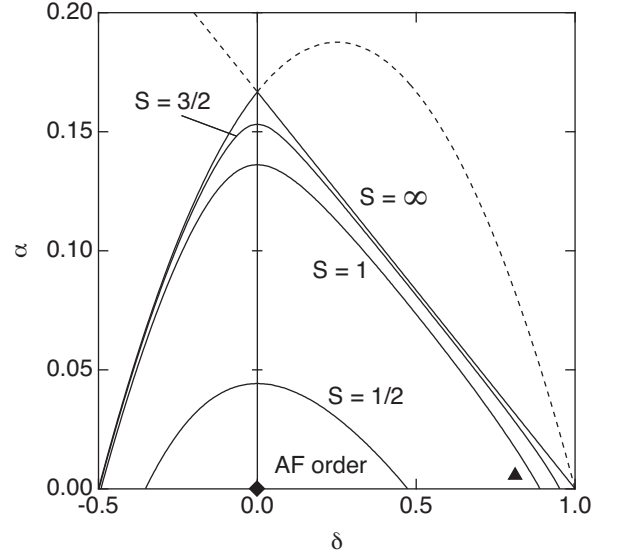}
\end{center}
\caption{Phase diagram 
in the space of distortion parameter $\delta$ 
and frustration parameter $\alpha$ (=$J_2/{\bar J}_1$). 
	Phase boundaries for $S$=$\frac{1}{2}$, 1, and 
$\frac{3}{2}$, and also the classical limit of the AF phase 
($S$=$\infty$) are shown. 
	For each $S$, the region below the boundary 
is an AF order phase, while that above 
is a disordered spin-gap phase. 
	Possible positions for InCu${}_{2/3}$V${}_{1/3}$O$_3$ 
and Na$_3$Cu$_2$SbO$_6$ are shown by 
diamond and triangle symbols respectively. 
} 
\label{phase}
\end{figure}

	The momentum cutoff $\Lambda$ is determined by 
keeping the number of degrees of freedom for ${\bf m}$. 
	The variable ${\bf m}$ is originally defined 
for each hexagonal block with sides of length $\sqrt{3}a$ 
(dash-dotted line in Fig.~\ref{phase}) 
and with area $\omega = (9\sqrt{3}/2)a^2$. 
	Then, in the continuum limit, the momentum cutoff 
$\Lambda$ is given as 
$a\Lambda = (2/3) (2\pi)^{1/2} 3^{-1/4}$ 
from $(2\pi)^2/\omega$ = $\pi \Lambda^2$. 

	We introduce rescaled dimensionless coordinates, 
$x_0 = \Lambda v_0 (c_+ c_-)^{\frac{1}{4}} \tau$, 
$x_1 = \Lambda (c_- / c_+)^{\frac{1}{4}} x$ and 
$x_2 = \Lambda (c_+ / c_-)^{\frac{1}{4}} y$ 
with $v_0 = 3 Sa{\bar J}_1/\sqrt{2}$. 
	The action (\ref{action_prime}) with neglecting 
the $A_Q$ term is then rewritten in a standard NLSM form: 
\begin{eqnarray}
A_{\rm eff} = \frac{1}{2g_0} \int \! d^3 x 
\left( \frac{\partial {\bf m}}{\partial x_{\mu}} 
\right)^2 
\label{action_rescaled} 
\end{eqnarray}
with coupling constant 
$g_0 = 3\sqrt{2}a\Lambda S^{-1}(c_+c_-)^{-\frac{1}{4}}$. 

      Applying the one-loop RG analysis 
by Chakravarty {\it et al.}~\cite{Chakravarty} to 
Eq.~(\ref{action_rescaled}), 
a quantum phase transition from the N\'eel phase with 
AF order to a disordered phase 
takes place at $g_0 = 4\pi$. 
	Rewriting this, the critical value $\alpha_c$ 
for the phase boundary 
in the $\delta$-$\alpha$ space is given by 
\begin{eqnarray}
\alpha_c = \frac{1}{6} \left[ 1 - \delta^2 
- \sqrt{\delta^2 ( 1 - \delta)^2 
+ \lambda_{S}^2} \, \right] 
\label{critical_general} 
\end{eqnarray}
with $\lambda_S = (\sqrt{3}\pi S^2)^{-1}$. 
	Parameter $\lambda_S$ is a measure of quantum character 
for the spin system. 
	In the classical limit of $\lambda_S = 0$ ($S=\infty$), 
 Eq.~(\ref{critical_general}) reduces to 
$\alpha_c = \frac{1}{6}(1-\delta)$ for $\delta > 0$ and 
$\alpha_c = \frac{1}{6}(1+\delta -2\delta^2)$ 
for $\delta < 0$; 
the present theory is applicable below the classical line. 
	The phase boundaries (\ref{critical_general}) 
for $S=\frac{1}{2}$, $S=1$ and $S=\frac{3}{2}$ are 
shown in Fig.~\ref{phase}. 
	For each $S$, the region above the solid line represented 
by the equation is a disordered phase with spin-gap and 
the region below is an AF ordered phase. 
	Possible positions for materials 
InCu${}_{2/3}$V${}_{1/3}$O$_3$ 
and Na$_3$T$_2$SbO$_6$ are also shown. 

	In the case of no distortion ($\delta$ = 0), 
Eq.~(\ref{critical_general}) becomes simply 
$\alpha_c$ = $\frac{1}{6}(1 - \lambda_S)$; e.~g. 
$\alpha_c =$ 0.044 ($S=\frac{1}{2}$), 0.136 ($S=1$), 
and 0.153 ($S=\frac{3}{2}$). 
	For $S=\frac{1}{2}$, the disordered phase has a range 
of 74\% against that of the classical AF order 
on the $\alpha$-axis (Fig.~\ref{phase}).  
	The actual portion may be fairly reduced from this, 
since the NLSM method includes, e.~g., 
a continuum  approximation~\cite{replacement}. 
	Whether or not a disordered phase survives for $\delta$ = 0 
will be considered in comparison 
with the square-lattice case. 
	The portion of the disordered range in the $J_1$-$J_2$ 
model on a square lattice is 
$(0.5 - 0.18)/0.5 \times 100 = 64$ \%  \cite{Takano}. 
	A disordered phase exists more plausibly for a honeycomb 
lattice than for a square lattice as was physically expected. 
	For $S \ge 1$, the disordered phase in Fig.~\ref{phase} 
is very narrow, 
and so its actual existence does not seem to be expected. 

 	In the limit of no frustration ($\alpha$ = 0), 
 Eq.~(\ref{critical_general}) is reduced to 
$2\delta^3 - 3\delta^2 + 1-\lambda_S^2 =0$. 
	The critical value of distortion for $S=\frac{1}{2}$ 
is $\delta = 0.47$. 
	The disordered state on the $\delta$-axis of 
Fig.~\ref{phase} is explained 
by the tendency of forming singlet pairs 
at strong $J'_1$-bonds. 

	Summarizing, we formulated an NLSM method for a spin 
system on a honeycomb lattice in the case that  
the exchange interaction includes both frustration 
by second-neighbor AF interactions and dimer-like 
distortion in first-neighbor AF interactions. 
	Applying the RG analysis by Chakravarty 
et al.~\cite{Chakravarty}, 
we have a ground-state phase diagram in a parameter space.  
	A disordered spin-gap phase is continuously 
extended from a strong frustration regime to a strong distortion 
regime, showing that the disordered ground states in both 
the regimes are essentially the same. 
	In the case of $S = \frac{1}{2}$, the spin-gap phase for 
the spin system on a honeycomb lattice is larger than that for 
the $J_1$-$J_2$ model on a square lattice, 
suggesting that a disordered state likely exists. 

	Material Na$_3$T$_2$SbO$_6$ with T=Cu 
($S=\frac{1}{2}$) 
has a strong dimer-like distortion and a spin-gap unlike 
the other similar materials with T=Ni ($S=1$) and T=Co 
($S=\frac{3}{2}$)~\cite{Miura}. 
	The distortion may be induced because it is in the spin-gap 
phase. 
	In fact, within the spin-gap phase, 
the energy of spins decreases and the lattice energy increases,  
as the dimer-like distortion increases; the balance determines 
an actual distortion. 
	On the other hand, within the AF phase, the dimer-like distortion 
does not reduce the spin energy, being consistent with 
little distortion in materials with T=Ni and T=Co. 
	To find a material with small distortion in a spin-gap phase, 
it is reasonable to try $S=\frac{1}{2}$ materials with strong 
elastic constants as well as appropriate second-neighbor AF 
interactions on honeycomb lattices. 

	I would like to thank Yoko Miura, Masatoshi Sato, 
Tamifusa Matsuura and Dai Hirashima for discussion. 
	This work is partially supported by the Grant-in-Aid
for  Scientific Research from the Ministry of Education,
Culture, Sports, Science and Technology of Japan.  



\begin{references}
\bibitem{fitting} 
They argue that 
a high-temperature expansion for a collection of singlet dimers 
fairly fits to the experimental susceptibility data.
\bibitem{Einarsson2} 
	Reference \cite{Einarsson} includes a derivation of 
an NLSM for a honeycomb lattice without distortion. 
	This pioneering work however does not keep the degrees 
of freedom of the original spin system in variable transformation; 
i.e. a spin variables are written as two new variables and 
the doubling is compensated by a regularization.
\bibitem{replacement} 
In literature, $S$ only in coupling constant $g_0$ of an NLSM 
is sometimes replaced by $\sqrt{S(S+1)}$. 
Following this, $\lambda_{\frac{1}{2}} \simeq$ 0.735 
for $S=\frac{1}{2}$changes to 0.245 and then 
the disordered region seems plausibly narrow. 
However we do not adapt the replacement, 
since it has not been justified.
\end{references}
\end{document}